\documentclass{aastex61}
\usepackage{natbib}
\usepackage[version=3]{mhchem} 
\usepackage{graphicx}
\usepackage{verbatim} 
\usepackage{soul}

\newcommand\aastex{AAS\TeX}

\received{03/05/19}
\revised{12/08/19}
\accepted{16/08/19}

\shorttitle{\aastex\ distribution of $^{26}Al$ in the forming Solar System}
\shortauthors{Pignatale et al.}

\begin{document}

\title{Fingerprints of the protosolar cloud collapse in the Solar System I: Distribution of presolar short-lived $^{26}$Al}

\correspondingauthor{Francesco C. Pignatale}
\email{pignatale@ipgp.fr}

\author[0000-0003-0902-7421]{Francesco C. Pignatale}
\affil{Mus\'eum national d'Histoire naturelle, Institut de Min\'eralogie, Physique des Mat\'eriaux et de Cosmochimie, D\'epartement Origines et Evolution, UMR 7590, CP52, 57 rue Cuvier, 75005, Paris, FRANCE
}
\affil{Universit\'e de Paris, Institut de Physique du Globe de Paris, CNRS, 1 rue Jussieu, 75005 Paris, FRANCE}

\author{Emmanuel Jacquet}
\affil{Mus\'eum national d'Histoire naturelle, Institut de Min\'eralogie, Physique des Mat\'eriaux et de Cosmochimie, D\'epartement Origines et Evolution, UMR 7590, CP52, 57 rue Cuvier, 75005, Paris, FRANCE
}
\author{Marc Chaussidon}
\affil{Universit\'e de Paris, Institut de Physique du Globe de Paris, CNRS, 1 rue Jussieu, 75005 Paris, FRANCE}

\author{S\'ebastien Charnoz}
\affil{Universit\'e de Paris, Institut de Physique du Globe de Paris, CNRS, 1 rue Jussieu, 75005 Paris, FRANCE}



\begin{abstract}
VERSION: Accepted ApJ DATE: \today

The short-lived radionuclide $^{26}\ce{Al}$  is widely used to determine the relative ages of chondrite components and  timescales of  physical and thermal events that attended the formation of the Solar System. However, an important assumption for using $^{26}\ce{Al}$  as a chronometer is its homogeneous distribution in the disk. 

Yet, the oldest components in chondrites, the Ca-Al-rich inclusions (CAIs), which are usually considered as time anchors for this chronometer, show evidence of $^{26}$Al/$^{27}$Al variations independent of radioactive decay. Since their formation epoch may have been contemporaneous with the collapse of the parent cloud that formed the disk, this suggests that $^{26}$Al was heteregeneously distributed in the cloud. We model the collapse of such an heterogeneous cloud, using two different $^{26}$Al distributions (monotonic and non-monotonic), and follow its re-distribution in the first condensates and bulk dust that populate the forming disk.

We find that CAIs inherit the $^{26}$Al/$^{27}$Al ratio of the matter infalling at the time of their formation, so that variations of $^{26}$Al/$^{27}$Al among primordial CAIs can be accounted for, independently of radioactive decay. The prevalence of a canonical ratio among them and its necessity for the differentiation of the first planetesimals suggest a (monotonic) scenario where $^{26}$Al sharply rose relatively close to the center of the protosolar cloud and essentially remained at a high level outward (rather than decreased since). As the $ ^{26}$Al abundance would be relatively homogeneous after cessation of infall, this would warrant the use of the Al-Mg chronometer from the formation of ``regular'' CAIs onward,  to chondrules and chondrite accretion.

\end{abstract}


\section{Introduction} 
\label{intro}

Chondrites are made of a mixture of components (Ca-Al-rich inclusions (CAIs), chondrules and matrix) with widely different  thermal histories, chemical composition and isotopic distribution \citep{2003TrGeo...1..143S}. Internal Al-Mg isochrons of individual CAIs suggest formation and processing during a period of $\sim$200 kyr  \citep{2012E&PSL.331...43M,2017GeCoA.201...65M,2017GeCoA.201..103U,2019E&PSL.511...25K}. Bulk CAI isochrons suggest that the formation of their precursors was restricted to a narrower time interval, 20-100 kyr \citep{2006ApJ...646L.159T,2008E&PSL.272..353J,2011ApJ...735L..37L}.

These timescales are consistent with the assembling time of a protoplanetary disk due to the collapse of its parent cloud \citep{2011ARA&A..49...67W}. The formation of the precursor blocks of chondrites could have thus started during the building phase of the solar protoplanetary disk when the Solar System's parent cloud was collapsing \citep{2005A&A...442..703H,2012M&PS...47...99Y,2018ApJ...867L..23P}. 

The composition of the cloud's gas and dust (in the interstellar medium) is the result of different chemical and nuclear  processes in previous generations of stars \citep{2011ARA&A..49...67W,goderis2016,2018PrPNP.102....1L}, and isotopic diversities detected in meteorites \citep{2007ApJ...655.1179T,2009Sci...324..374T} point to a heterogeneous  isotopic distribution within the different environments where chondrites, planets and other Solar System objects accreted. 

Among the freshest contributions to the protosolar cloud, the most popular in cosmochemistry is the short-lived  radionuclide aluminum-26, whose past presence is evidenced in meteoritic material by excesses in magnesium-26, the isotope into which it decays with a half-life of 0.7 Myr \citep{Nishiizumi2004}. Its highest initial abundances, in terms of $^{26}\ce{Al}/^{27}\ce{Al}$, has been measured in CAIs and amounts to $5.2\times10^{-5}$, the so-called canonical value (e.g. \citet{2012E&PSL.331...43M}).  While $^{26}$Al may be produced in different stellar environments such as supernovae or AGB stars \citep{2018PrPNP.102....1L}, a Wolf-Rayet type star is gaining more consensus as the main source of this isotope in the Solar System \citep{2017ApJ...851..147D}. The variable $^{26}\ce{Al}/^{27}\ce{Al}$ ratios observed in CAIs and chondrules have been widely used to infer time differences between the  formation events of these chondrites components \citep{2014E&PSL.390..318M,2017SciA....3E0407B}. However, it is not clear whether different $^{26}\ce{Al}/^{27}\ce{Al}$ ratios actually reflect time differences between components or heterogeneity in the distribution of $^{26}\ce{Al}$ in the  disk \citep{2012M&PS...47.1948K,2014E&PSL.390..318M}.  Heterogeneity in the  $^{26}\ce{Al}$ within the forming Solar System would, in fact, erase the possibility to use  $^{26}\ce{Al}/^{27}\ce{Al}$ as a chronometer.

  Such an heterogeneity is, in fact, suggested by discrepancies in the absolute \ce{U}-\ce{Pb} ages and relative \ce{Al}-\ce{Mg} ages of chondrules \citep{2011ApJ...735L..37L}, and correlations between nucleosynthetic anomalies such as  $^{54}\ce{Cr}$ and radiogenic excesses of $^{26}\ce{Mg}$ \citep{Van Kooten2016}.
 \citet{2017SciA....3E0407B} measured Pb-Pb ages of 22 chondrules and found that many of them are consistent with those of CAIs, thus, pointing to a contemporaneity between many chondrules and CAIs. This is in contrast with Al-Mg measurements showing mineral and bulk $^{26}\ce{Al}$-$^{26}\ce{Mg}$ isochrons consistent with an age gap of about 1.5~Myr between the CAIs and chondrules \citep{2009Sci...325..985V,2013M&PS...48.1383K,luu2015,2018GeCoA.227...19C}.

If Pb-Pb ages date correctly the last melting event of chondrules, these discrepancies would imply that chondrules actually formed from a reservoir depleted in $^{26}\ce{Al}$ compared to most CAIs. Among the CAIs themselves, some, such as the Fractionated and Unidentified Nuclear isotopic properties (FUN) CAIs, or PLAty hibonite Crystals (PLACs), show very low $^{26}\ce{Al}$ 
\citep[and reference therein]{1979ApJ...228L..93L,1995Metic..30..365M,2016GeCoA.189...70K,2017GeCoA.201....6P}, in spite of appearing primordial objects.  Whether they are the result of re-heating processes or vaporisation and recondensation within a  $^{26}\ce{Al}$ poor environment, or produced at a time close to the canonical CAIs but from a $^{26}\ce{Al}$-poor reservoir is still enigmatic \citep{1995Metic..30..365M,2014GeCoA.145..206K}.

\citet{2004ApJ...616.1265B,2008E&PSL.268..102B} investigated the evolution of a color field injected on a patch of a marginally gravitationally unstable disk, mimicking injection of an isotopic anomaly by a supernova, and showed homogenization  to occur within only a thousand years. However, no cloud infall onto the disk is considered, and beside the supernova-in-disk injection scenario, the aluminum-26 would first be located in the protostellar cloud collapsing into the protostar+disk system.  \citet{2011ApJ...733L..31M}, considering infall, found that, the largest population of refractory (formed above $T=1400$~K) present in the disk is produced  around the end of the infall. This suggests that $^{26}\ce{Al}$ was introduced in the disk contemporary to the disk formation and not in an already formed disk, and if it was not homogeneous, the heterogeneity could have been preserved until the end of the formation of refractory  \citep{2011ApJ...733L..31M}. This warrants the study of the distribution of $^{26}$Al in a disk forming from the collapse of an isotopically heterogeneous cloud, so as to find out to what extent this maps into an heterogeneous or homogeneous $^{26}$Al/$^{27}$Al ratio in the disk during and after infall.

To this end, we extend on our previous work \citep{2018ApJ...867L..23P} that itself was built on the models of \citet{2005A&A...442..703H} and \citet{2012M&PS...47...99Y}. We had found that a relatively low angular momentum, entailing infall on an initially compact disk, allowed extensive evaporation of presolar matter, and production of CAIs early on (within the first ~80 kyr for our run parameters, consistent with radiochronometric constraints), while later-infalling material may survive in relatively pristine form. Many CAIs would have been transported outward by the viscous expansion of the disk and ended up at large heliocentric distances, mixed in with less processed matter, accounting for the paradoxical mix of grains with diverse thermal histories observed in chondrites, in particular carbonaceous chondrites \citep{2018ApJ...867L..23P}.  In the same framework, we now study the isotopic composition of aluminum, following injection in the disk, in order to first verify the compatibility of  our model with  the observations  of  $^{26}\ce{Al}$,  and, second, to asses how they constrain the initial  heterogeneous/homogeneous  distribution of this isotope.

\section{Methods and models}
\label{methods}

We use the model described in \citet{2018ApJ...867L..23P} and \citet{Charnoz2019}. No changes are made in the basic physics of the code.  To recall briefly the implemented features, the infalling cloud is described by a spherical  isothermal shell \citep{1977ApJ...214..488S} that collapses  with a constant accretion rate while conserving angular momentum, leading to the source term on the disk given by \citet{2005A&A...442..703H}.
Cloud material falls onto the forming disk within the so called centrifugal radius, $R_c(t)$. As shown by \citet{2005A&A...442..703H}, \citet{2012M&PS...47...99Y} and  \citet{2018ApJ...867L..23P}, $R_c(t)$ corresponds to the location in the keplerian disk where the specific angular momentum equals that of the infalling cloud material; as the cloud collapses from inside out, it increases with time. In appendix~\ref{formule} we report the mathematical expressions of the accretion rates and $R_c(t)$.

 The code computes self-consistently grain growth, fragmentation and transport of dust particles and includes a  physics for  the dead zone, and simple chemical transformations \citep{2018ApJ...867L..23P}. Our chosen set of parameters for these simulations are: $T_{cd}=15$~K, $\Omega_{cd}=10^{-14}$ rad/s, $M_{0,\star}=0.02M_{\odot}$, $T_{\star}=4000$~K, $R_{\star}=3R_{\odot}$, $M_{tot}=1 M_{\odot}$, $\alpha_{active}=10^{-2}$, $\alpha_{dead}=10^{-5}$, $v_{frag}=10~{\rm m s^{-1}}$.

The  chemical composition of the cloud comprises \ce{H2}(g), \ce{H2O}(ice), silicates, iron, moderately volatiles species, and  refractories. Similarly to  \citet{2018ApJ...867L..23P}, we allow the infalling interstellar material to be either vaporised, ``processed'' (i.e. heated at temperatures $T > 800$~K but not vaporized) or left pristine upon its arrival in the disk.  In Appendix~\ref{composition}, we report our fiducial cloud abundances in Table~\ref{table1}, while Table~\ref{table2} summarizes all the implemented rules, with relative temperatures for each change of state. In addition to the ``background'' refractories above, we include $^{26}$Al.  We assume that all the considered  $^{26}$Al  is in a single separated refractory species. This is consistent with an injection of $^{26}$Al from an external independent source (for example a Wolf-Rayet star). We choose the refractories as a carrier because the $^{26}$Al produced by the star would condense into refractory as a first solid phase  \citep[]{1995GeCoA..59.3413Y,2011MNRAS.414.2386P} and anyway would follow the fate of common Al during thermal processing. Our $^{26}$Al-refractory is tagged as ``pristine'' in the cloud as it has not (yet) experienced any transformation in the disk forming process.

We assume the protosolar cloud to be zoned in $^{26}$Al, meaning that the proportion of $^{26}$Al in the infalling matter will depend on time, as the cloud collapses sequentially, from the inside out. As time elapses, $^{26}$Al is introduced in the forming disk with  early injection stages corresponding to the core of the cloud while later injection stages sample regions closer to the cloud's surface .
 \begin{figure}
 \begin{center}
{\includegraphics[width=0.5\columnwidth]{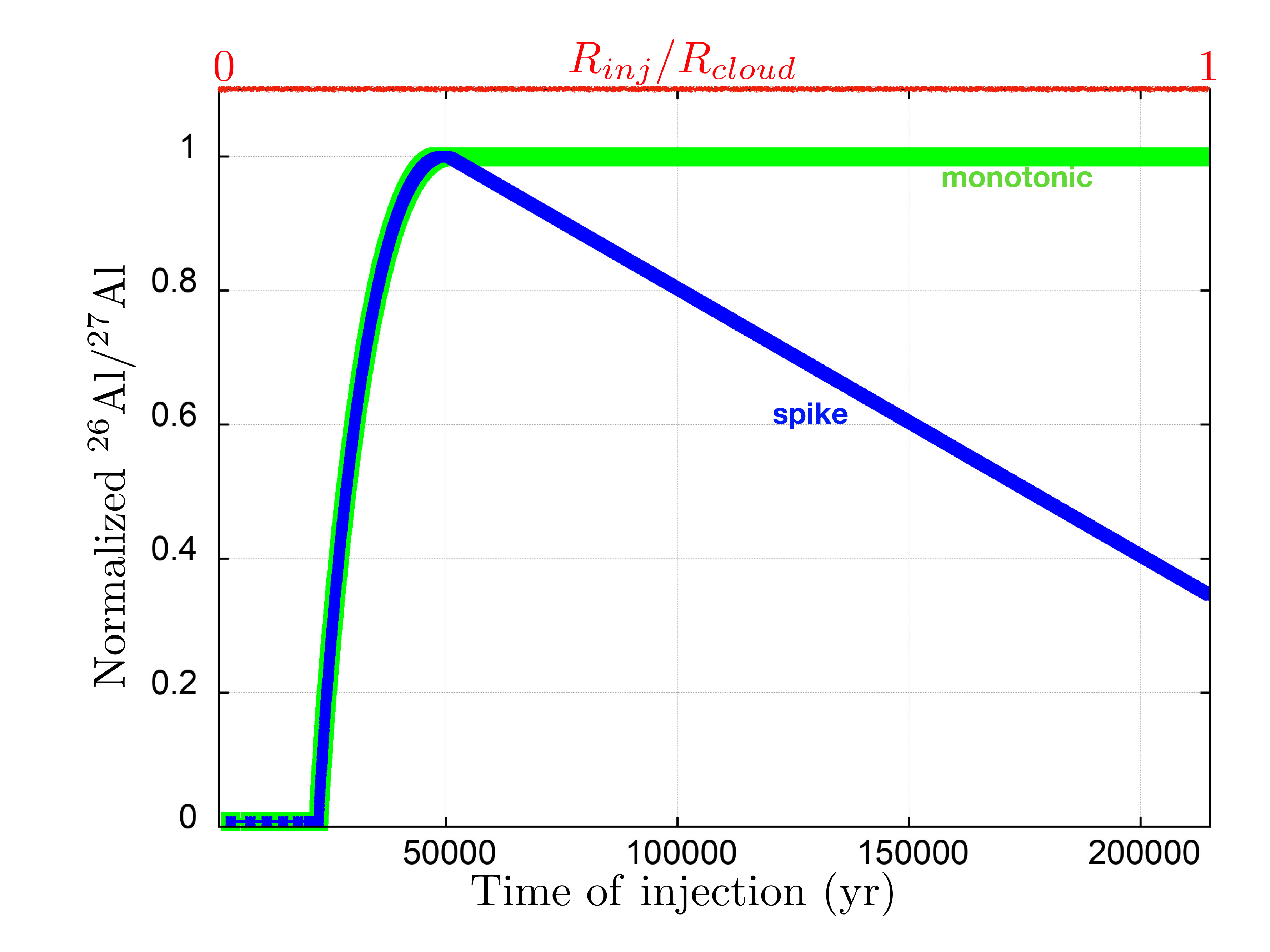}}
 \end{center}
\caption{Fiducial normalized $^{26}$Al  (i.e. $^{26}$Al/$^{27}$Al divided by the maximum value in the cloud) distributions in the parent cloud. The ``spike'' and ``monotonic'' functions are the same until they reach the maximum. They, then, diverge, with the former decreasing and the latter remaining constant.} 
\label{function}
\end{figure}

We present here the results assuming two arbitrary distributions of  $^{26}$Al  within the cloud  (Fig.\ref{function}), in terms of the  ``normalized $^{26}$Al'', i.e. the $^{26}$Al/$^{27}$Al ratio divided by the maximum value in the cloud. We call these ``injection functions''.
Although arbitrary, the shapes of these injection functions represent the two simplest possible scenarios imposed by the results in \citet{2018ApJ...867L..23P}, isotopic determination of CAIs age \citep{2012Sci...338..651C,2014E&PSL.390..318M} and the fact that most ``normal'' CAIs cluster around a  canonical $^{26}\ce{Al}/^{27}\ce{Al}$.
Following the results presented in \citet{2018ApJ...867L..23P}, in order to be actively transported by viscous expansion toward the outer disk, where carbonaceous chondrites are thought to form, our refractory condensates (hereafter CAIs, for ease of understanding) must form within the first 80 kyr of the  collapse, from presolar material that originates close to the center of the cloud \citep{2018ApJ...867L..23P}. The two considered distributions thus assume an increase from essentially zero to the maximum $^{26}$Al abundances within the first 80 kyr (similar to the sketch for  $^{26}\ce{Al}/^{27}\ce{Al}$ and $^{41}\ce{Ca}/^{40}\ce{Ca}$ suggested by \citet{1998ApJ...509L.137S}). The two injection functions, then, diverge, after the peak: one remains at the maximum value and will be called ``monotonic'' hereafter, while the other undergoes a  decrease, and will be referred to as the ``spike''. 

  The normalized $^{26}$Al abundances shown in this paper (including the aforementioned injection functions) will be \textit{decay-compensated}, i.e. multiplied by $e^{\lambda t}$ with $\lambda$ the decay constant of $^{26}$Al. This is because the use of $^{26}$Al as a chronometer amounts to assuming that this decay-compensated parameter is constant and uniform\footnote{This is incidentally a somewhat stronger requirement than simple spatial homogeneity in a given region, although of course if the disk is homogeneous throughout at some epoch, it will remain so at the same level hereafter.}. Indeed, stating that for two rocks A and B formed at times $t_A$ and $t_B$, their $^{26}$Al/$^{27}$Al ratios at their formation times are related by
  \begin{equation}
  \left(\frac{^{26}\mathrm{Al}}{^{27}\mathrm{Al}}\right)_B =  \left(\frac{^{26}\mathrm{Al}}{^{27}\mathrm{Al}}\right)_A e^{-\lambda (t_B-t_A)}
  \end{equation} 
  is equivalent to:
    \begin{equation}
  \left(\frac{^{26}\mathrm{Al}}{^{27}\mathrm{Al}}\right)_B e^{\lambda t_B}=  \left(\frac{^{26}\mathrm{Al}}{^{27}\mathrm{Al}}\right)_A e^{\lambda t_A}.
  \end{equation} 
  So this is the relevant quantity whose spatio-temporal variability must be assessed. For simplicity, we will hereafter drop the adjective ``decay-compensated'' from the phrase ``normalized $^{26}$Al abundance'' (with the understanding that the adjective ``normalized'' henceforth also refers to this operation).

\section{Results}
\label{results}

\begin{figure}
 \begin{center}
{\includegraphics[width=0.5\columnwidth]{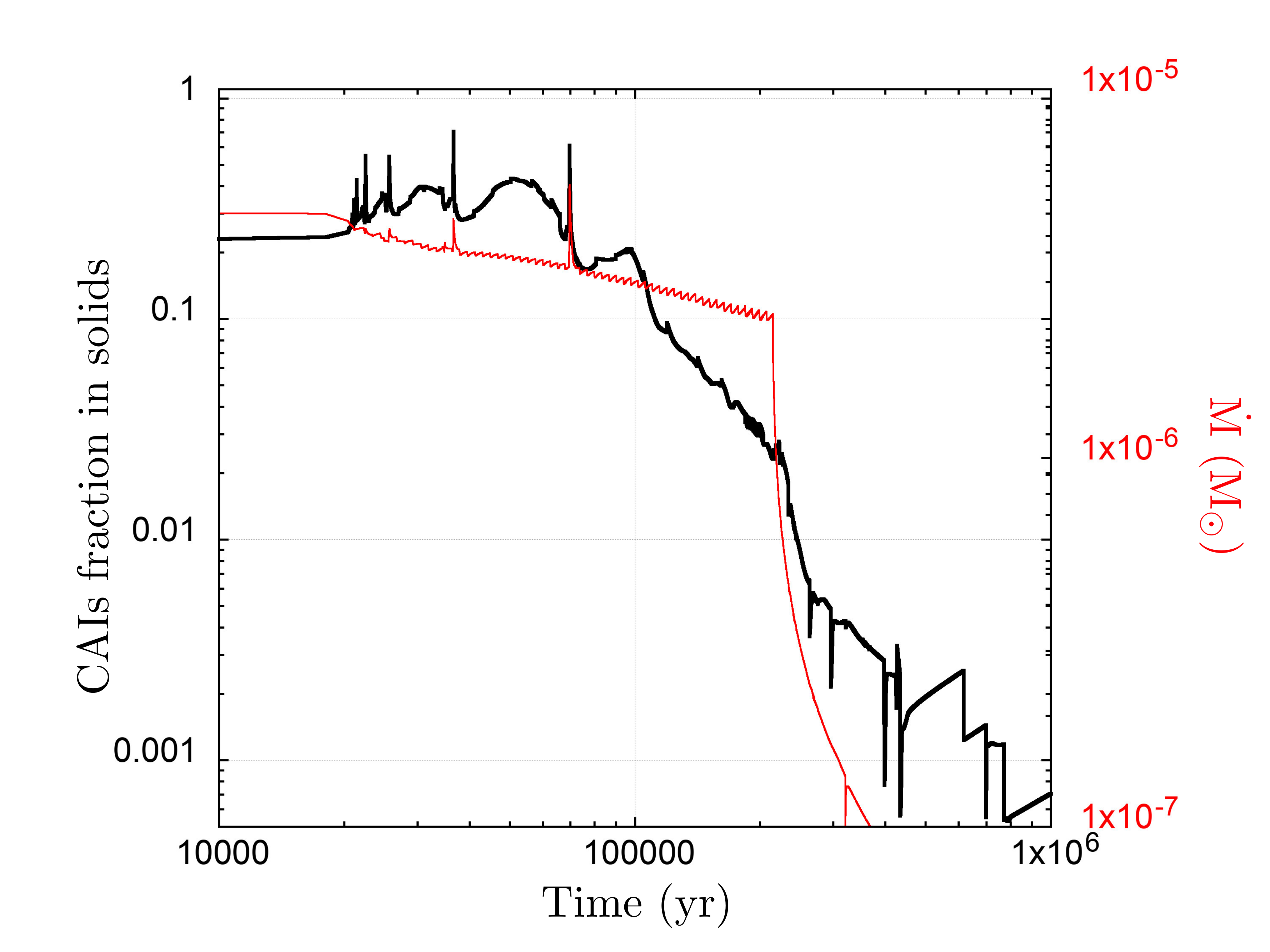}}
 \end{center}
\caption{Mass fraction of refractory condensates (CAIs) among the solids of the whole disk as a function of time and mass accretion rates. Peaks of CAIs production occur as a consequence of accretion bursts \citep{2018ApJ...867L..23P}. 
} 
\label{burst}
\end{figure}

In this section we focus on the time evolution of the distribution of the injected $^{26}$Al  in two different dust components: condensate refractories and bulk (i.e. condensates, processed and pristine  refractory dust). The time evolution of the mass of the star, disk and star+disk, condensation fronts, centrifugal radius, disk edge, dead zone,  surface densities of different species and mass fraction of rocky components at the end of the collapse, are shown in Appendix~\ref{diskbuilding}.
Our calculations are in agreement with the work of \citet{2012M&PS...47...99Y} and \citet{2018ApJ...867L..23P}.
In early times, all the presolar  refractories infalling at the highest temperature are vaporized and, as the gas is advected out, recondense as Solar System solids at the refractory condensation front.  As the centrifugal radius crosses the refractory condensation front it would not generally vaporise the injected  refractory dust that is coming from the cloud.  However, dust is still injected close enough and episodes of vaporisation and recondensation can occur, in particular during accretion bursts.  Figure~\ref{burst} shows the mass fraction of refractory condensates as a function of time, plotted together with the mass accretion rates (right y-axis).  Similarly to \citet{2018ApJ...867L..23P} we see that  peaks of CAIs production occur at each burst.

\begin{figure}[!htbp]
 \begin{center}
 {\includegraphics[width=0.45\columnwidth]{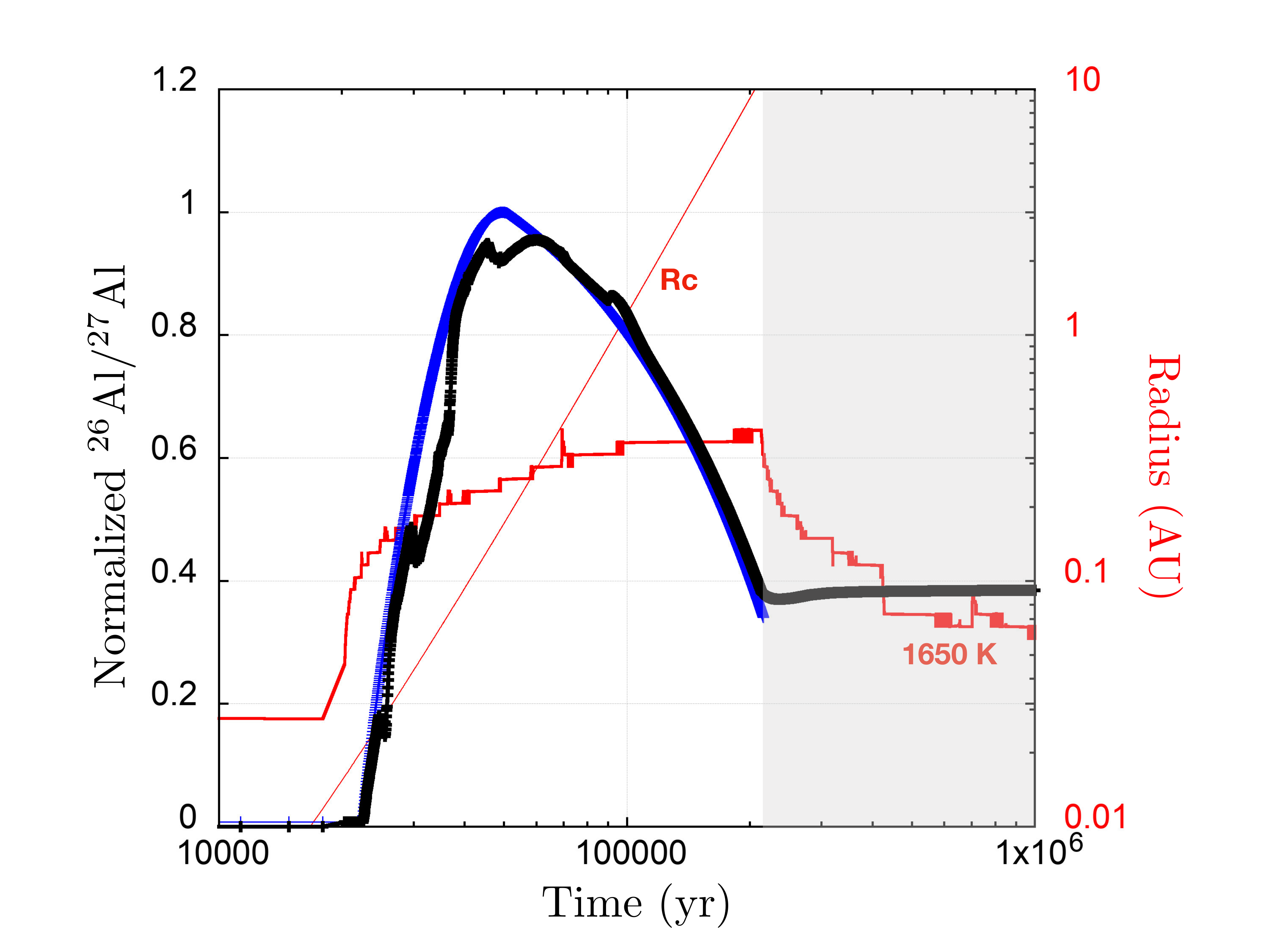}}
{\includegraphics[width=0.45\columnwidth]{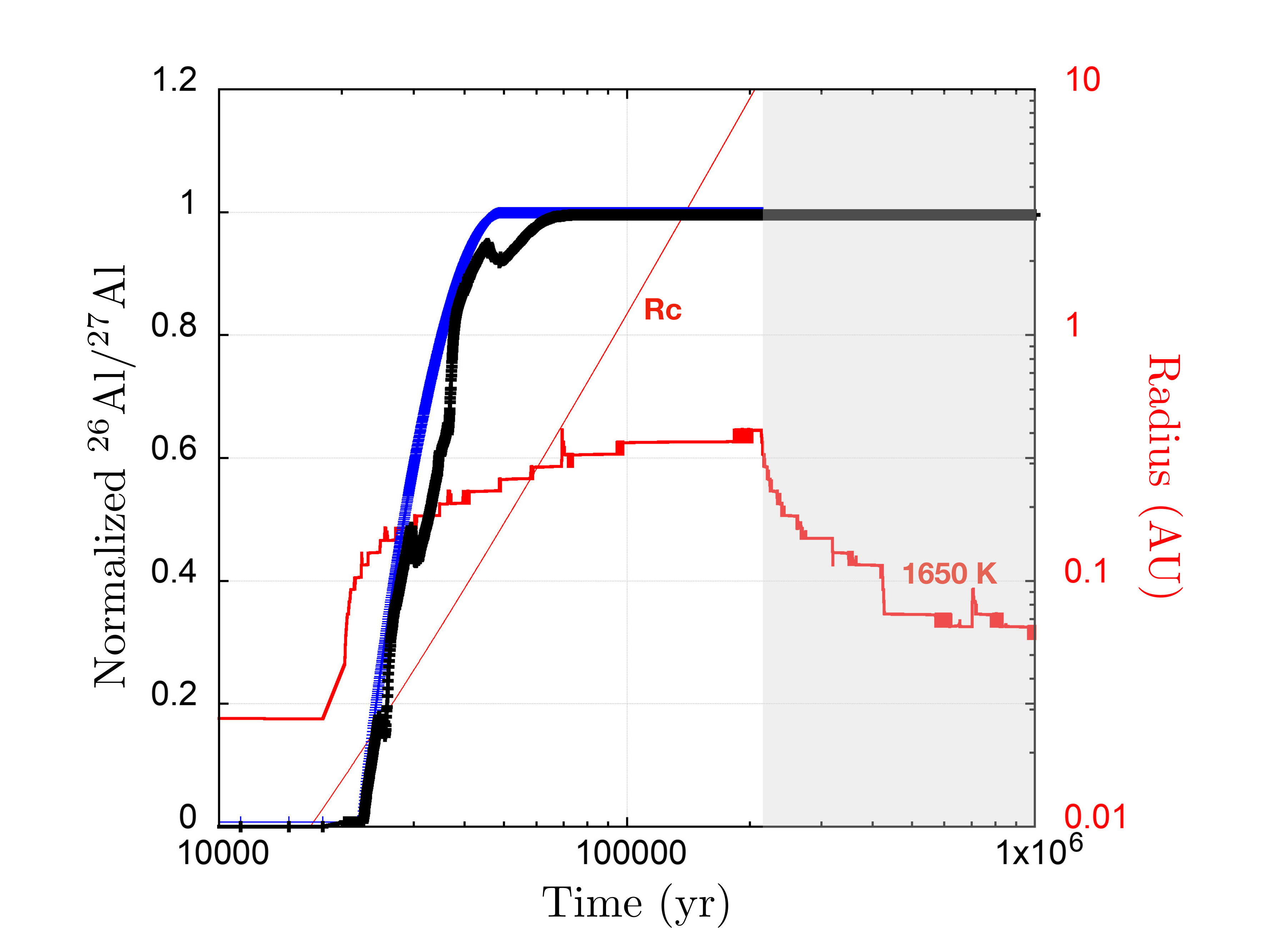}}
 \end{center}
\caption{Time evolution of the normalised $^{26}$Al abundance at the refractory condensation front (black `+' line) for the two chosen injection functions (blue `+' line): left, spike, and right, monotonic.  The ratio at the condensation front shows similarities with the cloud function. The grey-shaded area marks the epoch where infall has ceased. The time evolution of the locations of the condensation front and centrifugal radius are also shown, right y-axis).
} 
\label{confront}
\end{figure}

Figure~\ref{confront} shows the time evolution of the  normalized $^{26}$Al/$^{27}$Al ratio  (thick black crossed line) at the refractory condensation front (that is, the normalized $^{26}$Al abundance of CAIs as a function of their formation time), and simultaneously the ratio of the infalling matter (thick blue crossed line). Also shown are the time evolutions of the location of the refractory condensation front (red thick line, right y-axis) and centrifugal radius (thin red line, right  y-axis).
It is seen that the normalized abundance at the refractory condensation front closely follows that of the infalling matter, if with some lag, suggesting efficient transport from the centrifugal radius. The first CAIs produced are thus characterized by a low $^{26}$Al content. Later on, around $t\sim50$~kyr, the abundance reaches the maximum for the two functions and produces CAIs with higher $^{26}$Al. Note that for the spike, the maximum for CAIs (which would amount to the canonical value) would be slightly lower than the maximum in the cloud. The two functions then diverge, and the spike will again add CAIs with lower  $^{26}$Al, while the monotonic will keep injecting high contents of the radionuclide.  Moreover, bursts are also reflected into the shape of the $^{26}$Al/$^{27}$Al ratio at the refractory condensation front (Fig.~\ref{confront}).   At the end of the cloud collapse, around $t\sim215$~kyr, the disk becomes an accretion disk and the condensation front starts to move inwards. Dust that will continue to be processed at this condensation front will  be accreted to the Sun or incorporated into growing planetesimals.

\begin{figure}[!htbp]
 \begin{center}
 {\includegraphics[width=0.45\columnwidth]{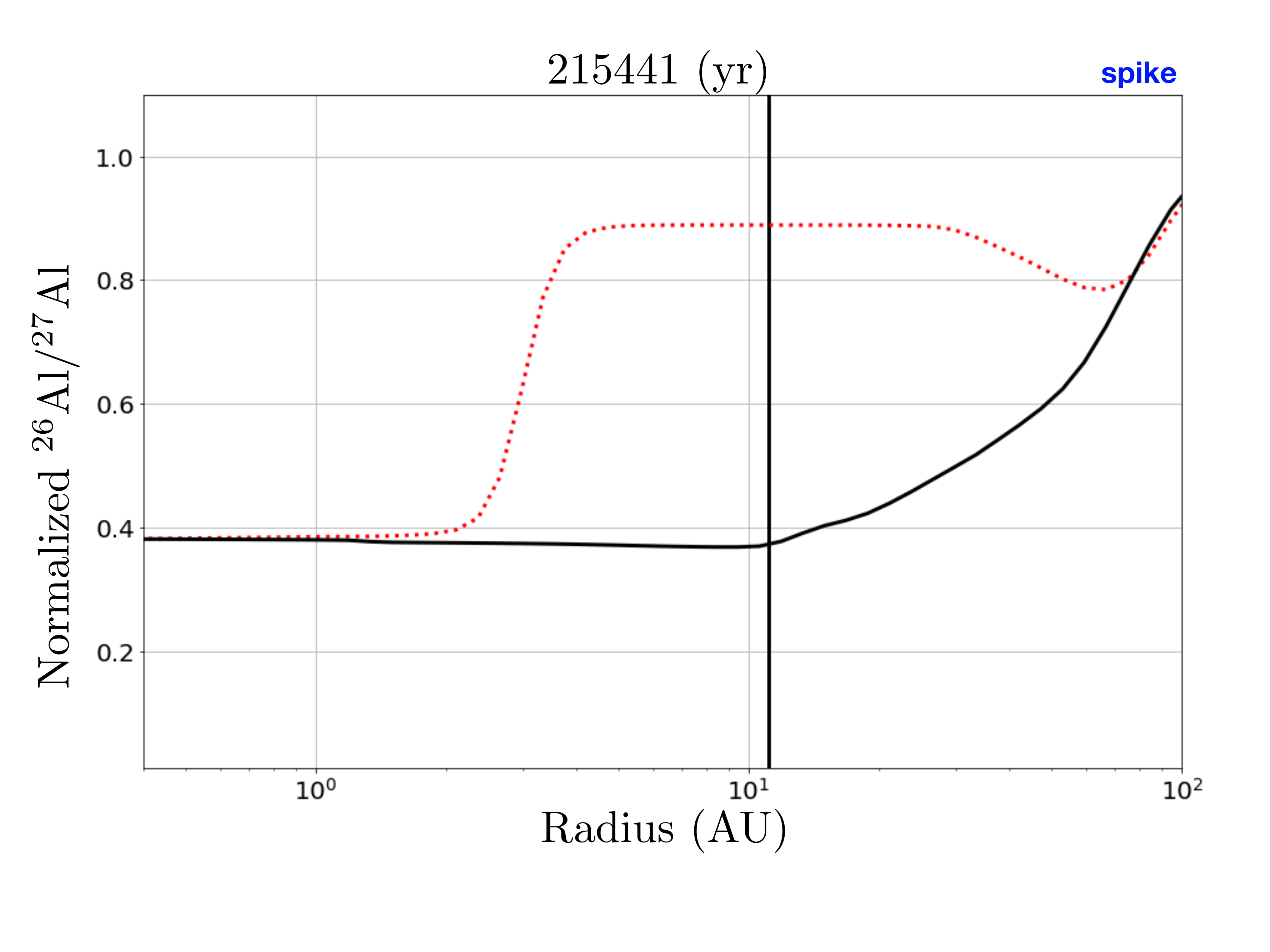}}
 {\includegraphics[width=0.45\columnwidth]{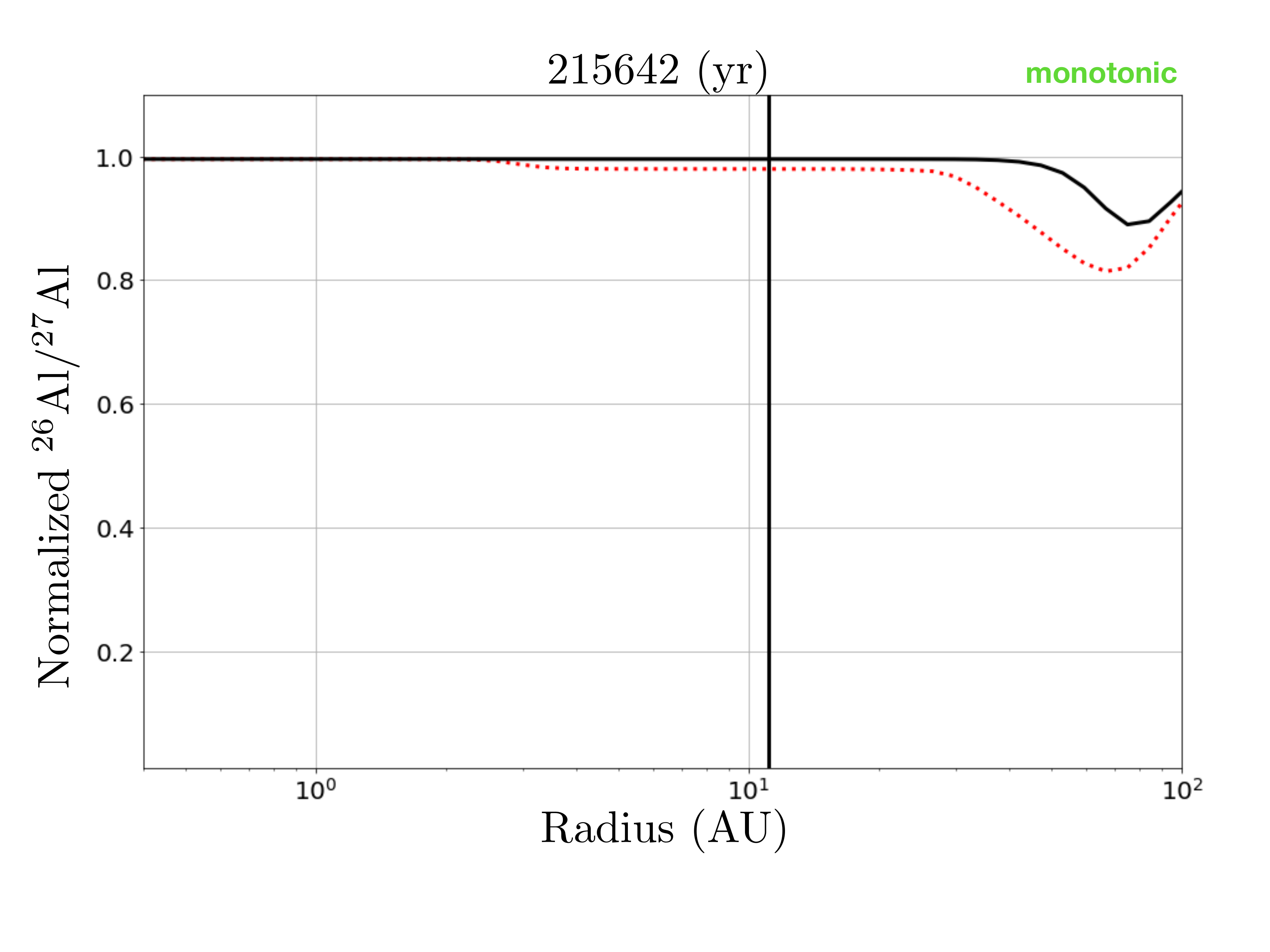}}\\
  {\includegraphics[width=0.45\columnwidth]{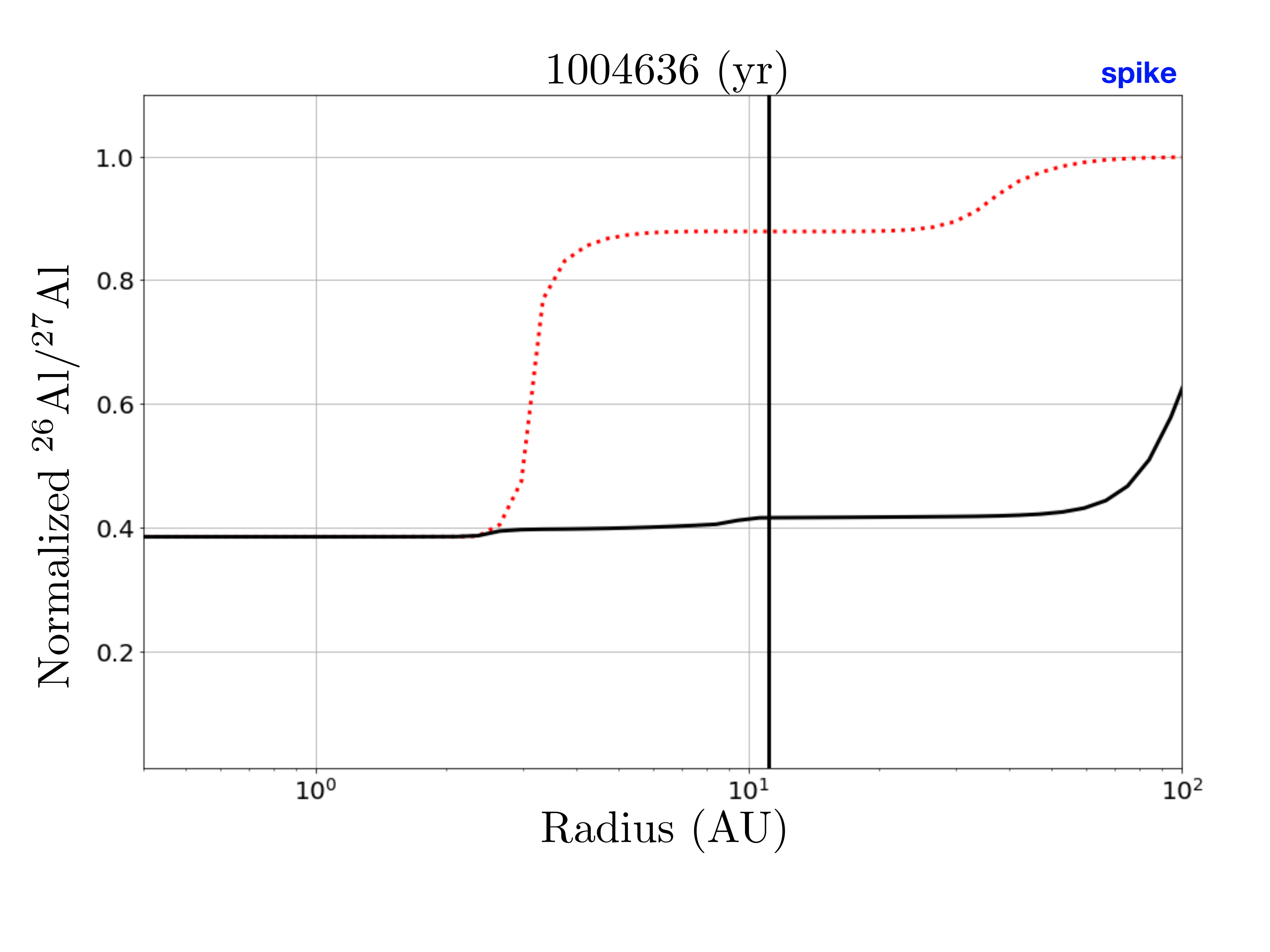}}
 {\includegraphics[width=0.45\columnwidth]{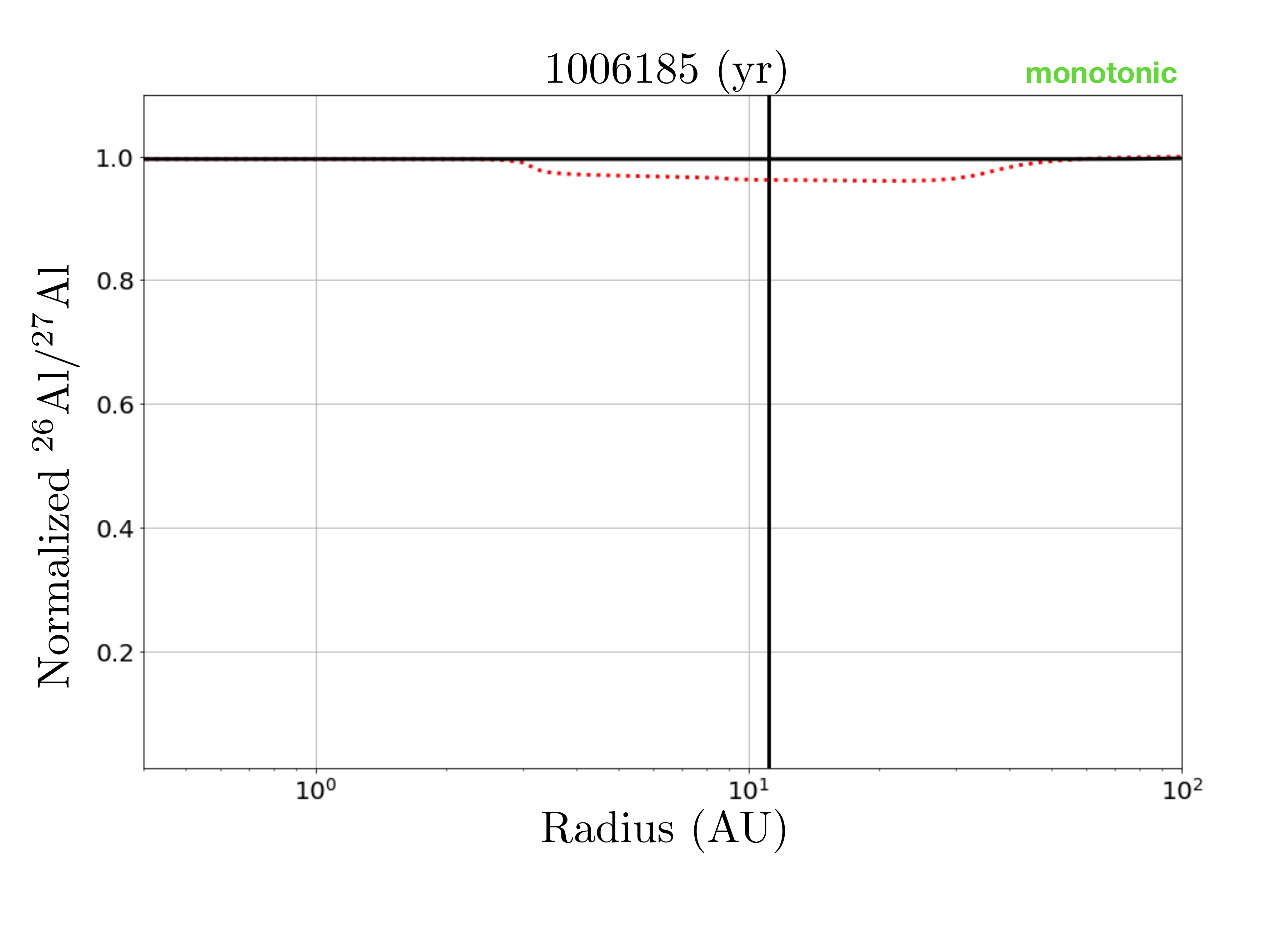}}\\
 \end{center}
\caption{Normalised $^{26}$Al abundance in CAIs (red dotted line) and bulk (black continuous line) at the end of collapse (top) and after 1~Myr (bottom), as a function of the heliocentric distance, for the spike (left) and monotonic (right) function.  The spike produces evident differences between CAIs and bulk in the outer disk ($R>3$~AU), and between CAIs in the inner and outer disk. The monotonic case produces instead similar values in all disk regions. The average radial distribution does not substantially change with time.
} 
\label{average}
\end{figure}

Figure~\ref{average} shows the  radial  distribution of the normalised $^{26}$Al/$^{27}$Al ratio in the CAIs (red dotted line) and in the bulk (black solid line) at the end of the collapse and after 1~Myr, for the two functions (spike left, monotonic right). Note that the value for the CAIs is only an average at a given heliocentric distance; the Eulerian nature of the code does not allow the actual distribution of individual inclusion within CAIs  to be given. In the case of the spike two different families of CAIs can be retrieved. The plateau where $R>3$~AU is characterized by the CAIs that experienced strong outward advection, after an early formation.
Inner disk CAIs ($R<3$~AU) are those that represent the local conditions (late production and reprocessing of dust). The transition radius $\sim 3$ AU corresponds to a break in the refractory condensate abundance profile (see Fig.\ref{diskbuild}c) which marks the previous expansion of the disk. The spike results in stark differences between the CAIs and bulk that are located in  the outer disk  ($R>3$~AU) because the bulk starts to include the $^{26}$Al-poorer processed and  pristine dust that is directly injected in the disk from the cloud, while CAIs may fossilize the conditions when they were produced most (Fig. \ref{confront}). Differences can be also clearly seen  between CAIs in the outer and inner disk. In the monotonic case, differences  between CAIs and bulk and between inner and outer CAIs are, instead, subtle because of the prolonged plateau of the injection function. If the latter has risen later, the difference would have been more significant; at the other extreme, of course, if the injection had been constant all the way, then all the curves would coincide with its level. These distributions are preserved with time (see Fig.~\ref{average}). So after infall, most matter has the same bulk $^{26}$Al normalized abundance (around 0.4 for the spike and 1 for the monotonic case).

\section{Discussion}
\label{discussion}
In its two studied varieties (monotonic or spike),  heterogeneous  injection can explain varying $^{26}$Al/$^{27}$Al in CAIs unrelated to radioactive decay (see Fig. \ref{confront}). Note that the variations would be essentially temporal  (but controlled by the gradient in the cloud rather than  by decay) rather than spatial. Thus, in both varieties of the injection function, FUN CAIs and PLACs could represent an early generation of CAIs before the canonical abundance was reached. This is suggested by the CAIs that are produced at the earliest stage of the simulations when the injection functions of the normalized $^{26}$Al/$^{27}$Al raise from zero to  canonical, $t\leq50$~kyr (see Fig. \ref{confront}), although the Eulerian code does not allow to track their exact proportion in subsequent times. Since this would be a temporal variation, there is no need for any supercanonical ratio elsewhere in the Solar System the lack of which appeared an objection to the inherited heterogeneity scenario  to \citet{2016AREPS..44..709D}. The (decay-compensated) $^{26}$Al is essentially uniform and constant after cessation of infall, meaning $^{26}$Al can be used as a relative chronometer in this period. 

The question is then whether this $^{26}$Al level would be the same as that of regular CAIs with canonical $^{26}$Al abundances (monotonic case), or lower (spike) as suggested by \citet{2017SciA....3E0407B} to reconcile Al-Mg and Pb-Pb dating. Only in the former case can the initial $^{26}$Al/$^{27}$Al ratios be used to infer age differences between regular CAIs and chondrules (assumed to be produced from the melting of ``bulk'' material). The spike function would produce at least three types of CAIs: an initially older $^{26}$Al-poor family followed by $^{26}$Al-higher  CAIs and, then, a younger $^{26}$Al-poorer population. However, the latter would hardly resemble FUN CAIs as they should essentially have converged toward the present-day composition for stable isotopes, at variance with the nucleosynthetic anomalies shown by FUN CAIs.

Also, the significant clustering of $^{26}$Al/$^{27}$Al ratios for mineral isochrons in CAIs \citep{2012E&PSL.331...43M} around the canonical value is an indication against a late CAIs population with significantly different $^{26}$Al  (as predicted  by the spike function), as they would tend to be preferentially preserved (\citet{2011ApJ...733L..31M}; and also our Figs. \ref{confront} and~\ref{average}), whereas the FUN CAIs mentioned above are a minor population \citep{2014GeCoA.145..206K}. 

In addition, bulk chondrites (which would have inherited the normalized abundance at the end of the infall) define a   $^{26}$Al isochron consistent with a canonical $^{26}$Al abundance despite large errors due to the small range of Al/Mg ratios 
 ($(6.3\pm 1.3)\cdot 10^{-5}$ \citep{2010E&PSL.297..165S} or $(4.7\pm 0.7) \cdot 10^{-5}$ \citep{2019E&PSL.522..166L}).

Thus, the $^{26}$Al in bulk chondrites is only in agreement with the results produced by monotonic function. Moreover, such an abundance is necessary to explain the differentiation of planetesimals over 2 Myr \citep {2005ASPC..341..915S} as suggested by Hf-W dating of iron meteorites \citep{2014E&PSL.403..317K} or HED achondrites \citep{2011SSRv..163..141M}. Interestingly, Al-rich chondrules, which are believed to have inherited their excess Al from precursor CAIs \citep{2002M&PS...37...91K,2017M&PS...52.2672J}, exhibit $^{26}$Al/$^{27}$Al ratios comparable to their mainstream ferromagnesian counterparts \citep{2009GeCoA..73.5080H}, around 3-15 times lower than the canonical values. This could not be the case if they had formed in the same epochs these CAIs condensed (provided they were representative of the CAIs found elsewhere in the same chondrites in terms of $^{26}$Al abundances), so the time difference must be real for them.

\section{Conclusions}
\label{conclusions}

In this work we investigated how different distributions of $^{26}$Al within the protosolar cloud  translate in the forming disk and how $^{26}$Al  would be incorporated in the earliest refractory condensates (CAIs) and bulk material (condensates plus processed and pristine refractory dust). CAIs would fossilise the evolution of the isotopic composition of the condensation front during infall. We find that low initial levels of $^{26}$Al (in the inner regions of the parental cloud) can account for FUN CAIs, before it rose to canonical value. We investigated two possibilities for the subsequent evolution of the aluminum-26 injection pattern: a plateau (the ``monotonic'' function) and a return to low values (the ``spike''). The spike would predict a significant population of $^{26}$Al-poor CAIs and little energy for planetesimal differentiation, unlike the monotonic one which seems  more consistent with observations. 

Thus, in the earliest infall stage, aluminum-26 cannot be used as a chronometer, at least in terms of the exponential decay law, but low $^{26}$Al/$^{27}$Al in a CAI  (if not ascribable to late resetting) could still be used to suggest early formation (e.g. if supported by nucleosynthetic anomalies). The $^{26}$Al relative chronometry would however retain validity from the formation of regular CAIs onward.

Our formalism can be also extended to any nucleosynthetic anomaly heterogeneously distributed in the protosolar cloud. This is the subject of a companion work \citep{Jacquet 2019}.

\acknowledgments
The authors wish to acknowledge the financial support of ANR-15-CE31-0004-1 (ANR CRADLE). We wish to thank M. Gounelle and A. Krot for useful discussions. We thank the anonymous referee for providing suggestions, references and comments that greatly improved the manuscript.

\newpage
\appendix

\section{Accretion rates and centrifugal radius}
\label{formule}

The cloud collapses onto the protostar-disk system with a constant accretion rate  given by \citet{1977ApJ...214..488S} (see also \citet{2012M&PS...47...99Y}: 
\begin{equation}
 \dot{M}= 0.975\frac{C^{3}_{cd}}{G} .\
\label{eq1}
\end{equation}
 In this equation $G$ is the gravitational constant,  $C_{cd}$ is the isothermal sound speed, $C_{cd}^{2}=k_{b}T_{cd}/\mu m_{p}$.  $k_{b}$ is the Boltzmann constant, $T_{cd}$ is the temperature of the cloud and $\mu=2.2$ is the molecular weight of the gas in terms of the proton mass, $m_{p}$.

Assuming angular momentum conservation, the surface density accreted below the centrifugal radius, $R_c(t)$, is \citep{2005A&A...442..703H,2012M&PS...47...99Y}:

\begin{equation}
\dot{\sigma}(r,t)= \frac{\dot{M}}{8\pi R^{2}_{c}} \Bigg( \frac{r}{R_{c}(t)} \Bigg)^{-3/2}  \Bigg[  1 - \bigg(\frac{r}{R_{c}(t)} \Bigg)^{1/2} \Bigg]^{-1/2} ,\
\label{eq2}
\end{equation}
where the expression of the centrifugal radius $R_c(t)$ is 
\begin{equation}
R_c(t)= 53\: \mathrm{AU}\Bigg( \frac{\omega_{cd}}{10^{-14} s^{-1}}  \Bigg)^{2}   \Bigg( \frac{T_{cd}}{10~K}  \Bigg) ^{-4}     \Bigg( \frac{M(t)}{1 M_{\odot}}  \Bigg) ^{3}   .\
\label{eq23}
\end{equation}
$\omega_{cd}$  is the constant angular velocity of the cloud, $T_{cd}$ is the temperature of the cloud and $M(t)$ is the total mass of the protostar+disk system at time $t$.

\section{Initial Cloud Composition and rules for dust transformation}
\label{composition}

Table~\ref{table1} reports the fiducial initial dust composition in the cloud. Values are retrieved using thermodynamic equilibrium with the method described in \citet{2011MNRAS.414.2386P}  using the Sun's elements abundances reported in \citet{2009ARA&A..47..481A} and with a closing temperature of $T=700$~K, i.e. before the conversion of \ce{CO}(g) to  \ce{CH4}(g). These values show very good agreement with those derived in \citet{2010fee..book..157L} assuming non-equilibrium at low temperatures. This values update the complete equilibrium values ($T=150$~K), used in \citet{2018ApJ...867L..23P}.

Table~\ref{table2} reports the simple thermodynamic rules implemented in the code. Differences from \citet{2018ApJ...867L..23P} include: (i) the introduction of moderately volatiles species (movo), (ii) the change of the processing temperature from $T=900$~K to $T=800$~K (for consistency with literature; e.g. \citet{2014Icar..237...84H}), (iii) the change of the condensation temperatures so as to match the thermodynamic calculation of \citet{2011MNRAS.414.2386P}. Moreover, we made two simplifications compared to our previous work:  (i) we removed the special rule for which metallic iron becomes processed iron at  $T=650$~K, and (ii) all the refractory pristine material that is not vaporised and falls between $800<T(\rm K)<1650$ is now considered processed refractory. See \citet{2018ApJ...867L..23P} for more details on the old rules.

\begin{table*}[!htbp]
\begin{center}
\begin{tabular}{|c|c|}
\hline 
species & mass fraction \\
\hline 
refractory (cloud) & 0.0958 \\
\hline 
refractory (anomaly) & varying\\
\hline 
silicates & 0.3191 \\
\hline 
iron & 0.1738 \\
\hline 
movo & 0.0458 \\
\hline 
\ce{H2O}(ice) & 0.1925 \\
\hline 
\ce{CO}(ice) & 0.7435 \\
\hline 
\ce{H2} & 98.4295 \\
 \hline 
\end{tabular} 
\caption{Fiducial initial dust composition in the cloud.}
\label{table1}
\end{center}
\end{table*}

\begin{table*}[!htbp]
\begin{center}
\begin{tabular}{|c|c|c|c|c|}
\hline 
species & T (K) & become & T  (K) & become \\ 
\hline 
\hline
pristine refractories &  $>1650$  & refractory(g) & &  \\ 
\hline 
refractory(g)  &   $<1650$ & CAIs$^{a}$ & $>1650$   &   refractory(g)  \\ 
\hline 
pristine refractories  &   $>800$ & processsed refractories  & $>1650$   &   refractory(g)  \\ 
\hline 
pristine refractories  &   $<800$ & pristine refractories  &    &    \\ 
\hline 
\hline
pristine silicates &  $>1415$  & silicates(g) & &  \\ 
\hline 
silicates(g)  &   $<1415$ & condensed silicates$^{a}$ & $>1415$   &   silicates(g)  \\ 
\hline 
pristine silicates  &   $>800$ & processsed silicates  & $>1415$   &   silicates(g)  \\ 
\hline 
pristine silicates  &   $<800$ & pristine silicates  &    &   \\ 
\hline 
\hline
pristine iron &  $>1450$  & iron(g) & &  \\ 
\hline 
iron(g)  &   $<1450$ & metallic iron$^{a}$ & $>1450$   &   iron(g)  \\ 
\hline 
pristine iron  &   $>800$ & processsed iron  & $>1450$   &   iron(g)  \\ 
\hline 
pristine iron  &   $<800$ & pristine iron  &    &  \\ 
\hline 
\hline
pristine movo &  $>800$  & movo(g) & &  \\ 
\hline 
movo(g)  &   $<800$ & condensed movo & $>800$   &   movo(g)  \\ 
\hline 
pristine movo  &   $>800$ &  movo(g)  &   &    \\ 
\hline 
\hline
water ice  & $> 150 $ &  water  vapour & $< 150 $   & water  ice   \\ 
\hline 
\hline 
CO ice & $> 25 $ &  CO(g) & $< 25 $ & CO ice  \\ 
\hline 
\hline
\ce{H2}(g) &  &   &  &  \\ 
\hline 
\end{tabular} 
\caption{Rules implemented in the code. We assume that the kinetics timescales of evaporation, condensation and processing are  instantaneous. (a) Same as in \citet{2018ApJ...867L..23P}, in our calculations condensed CAIs, condensed silicates and metallic iron, do not equilibrate with the surrounding environment as evidenced by actual chondrites.}
\label{table2}
\end{center}
\end{table*}

\section{Disk building and evolution}
\label{diskbuilding}
 \begin{figure}[!htbp]
 \begin{center}
{\includegraphics[width=0.45\columnwidth]{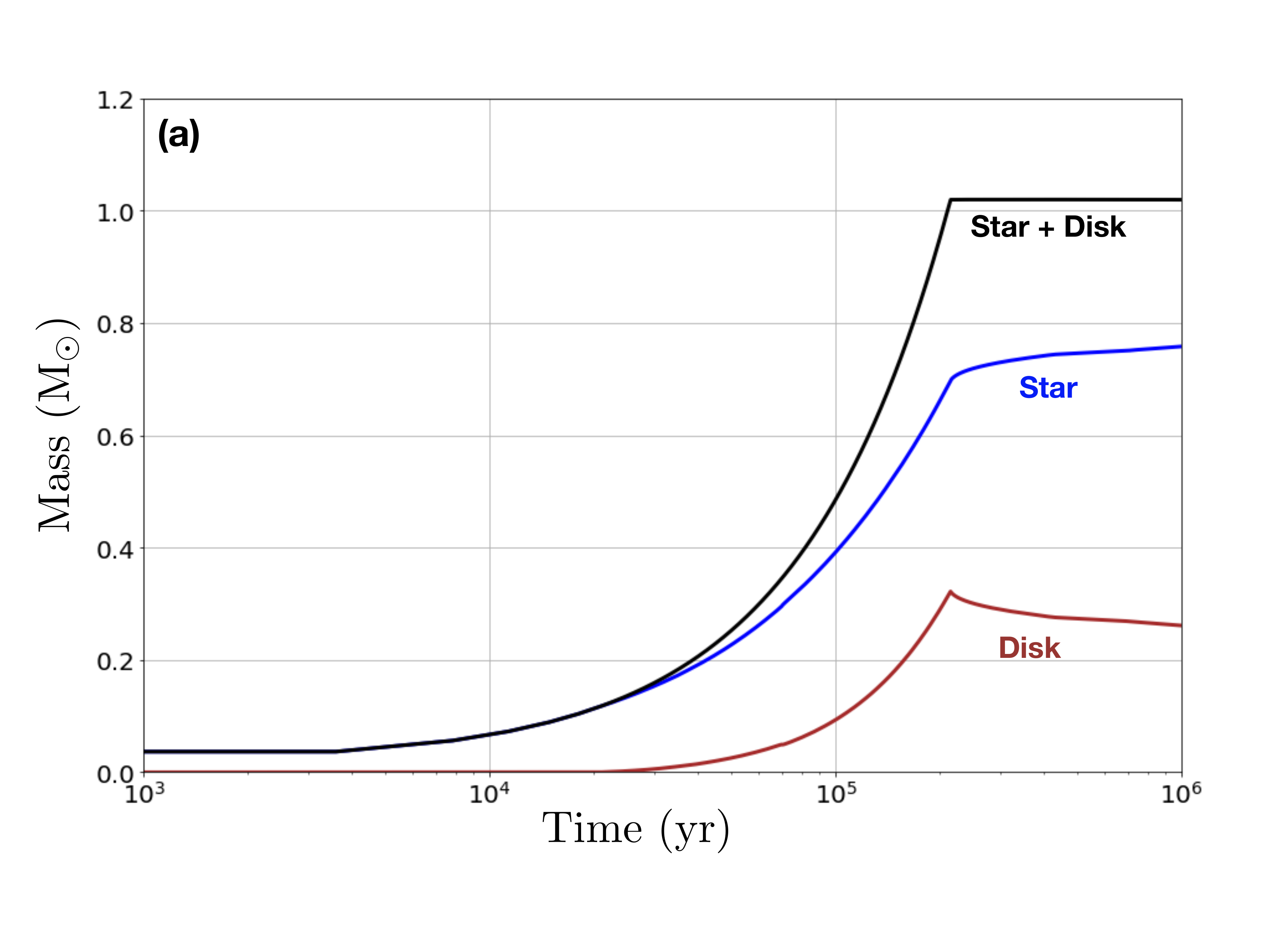}}
{\includegraphics[width=0.45\columnwidth]{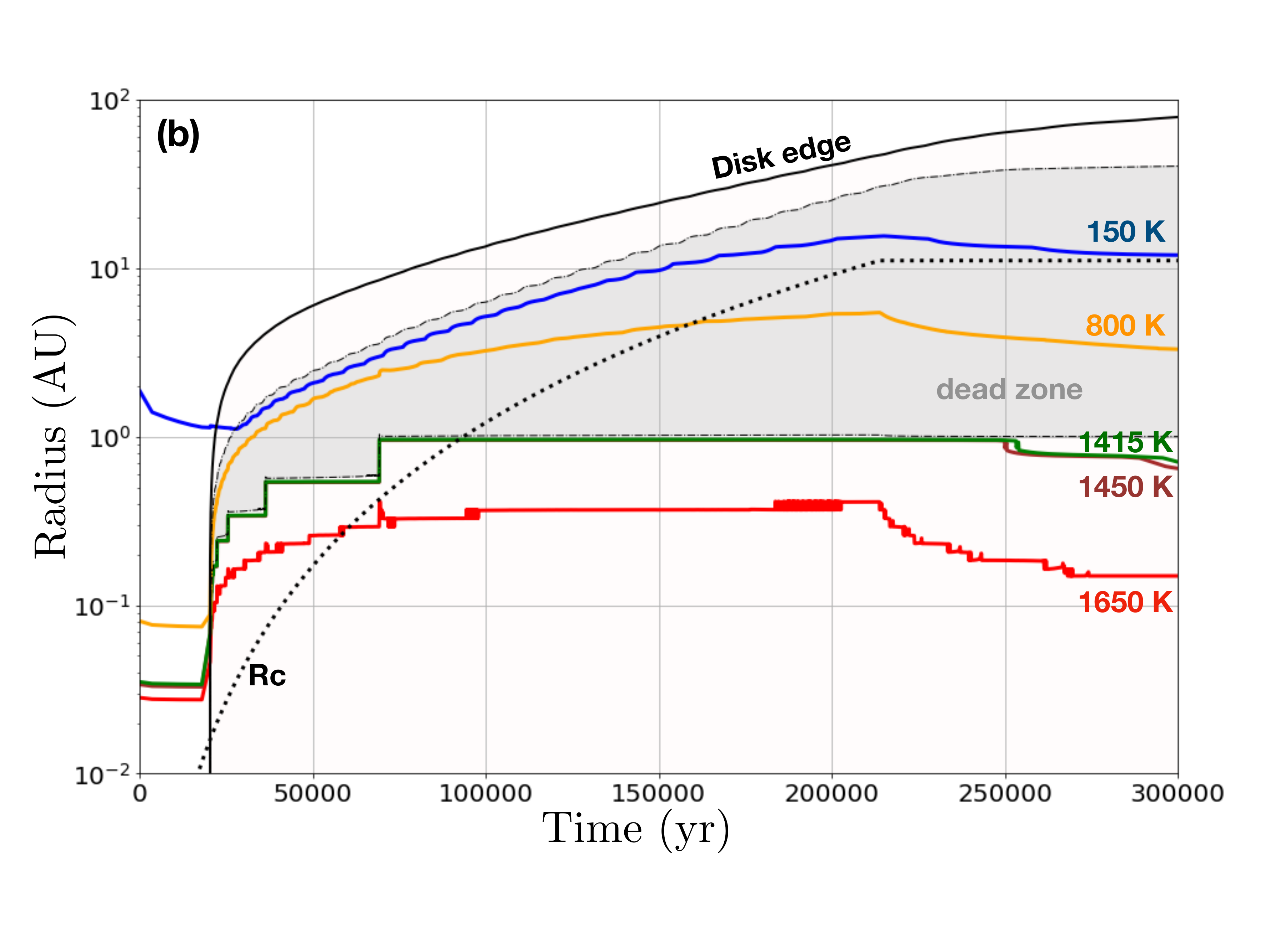}}\\
{\includegraphics[width=0.45\columnwidth]{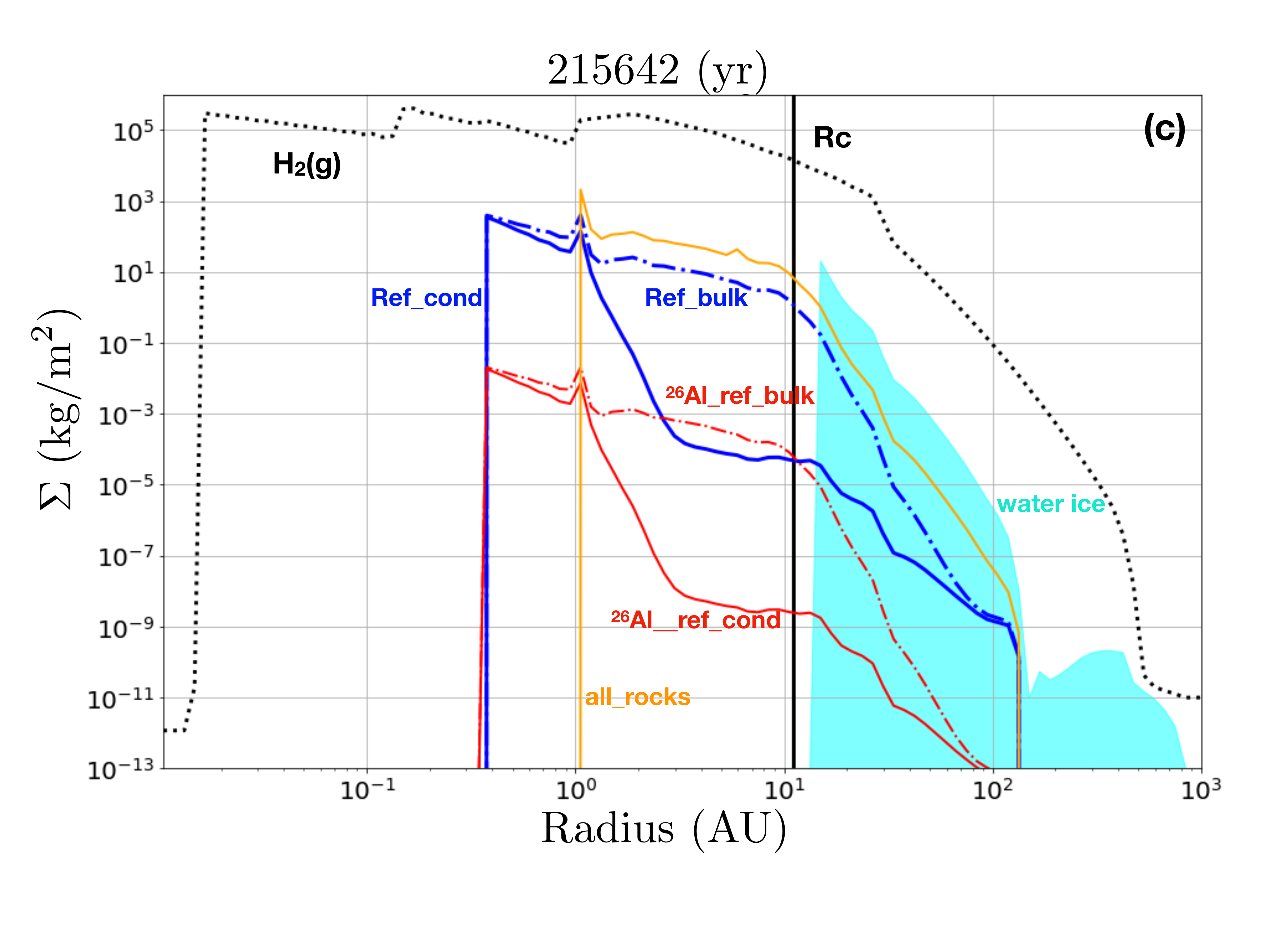}}
{\includegraphics[width=0.45\columnwidth]{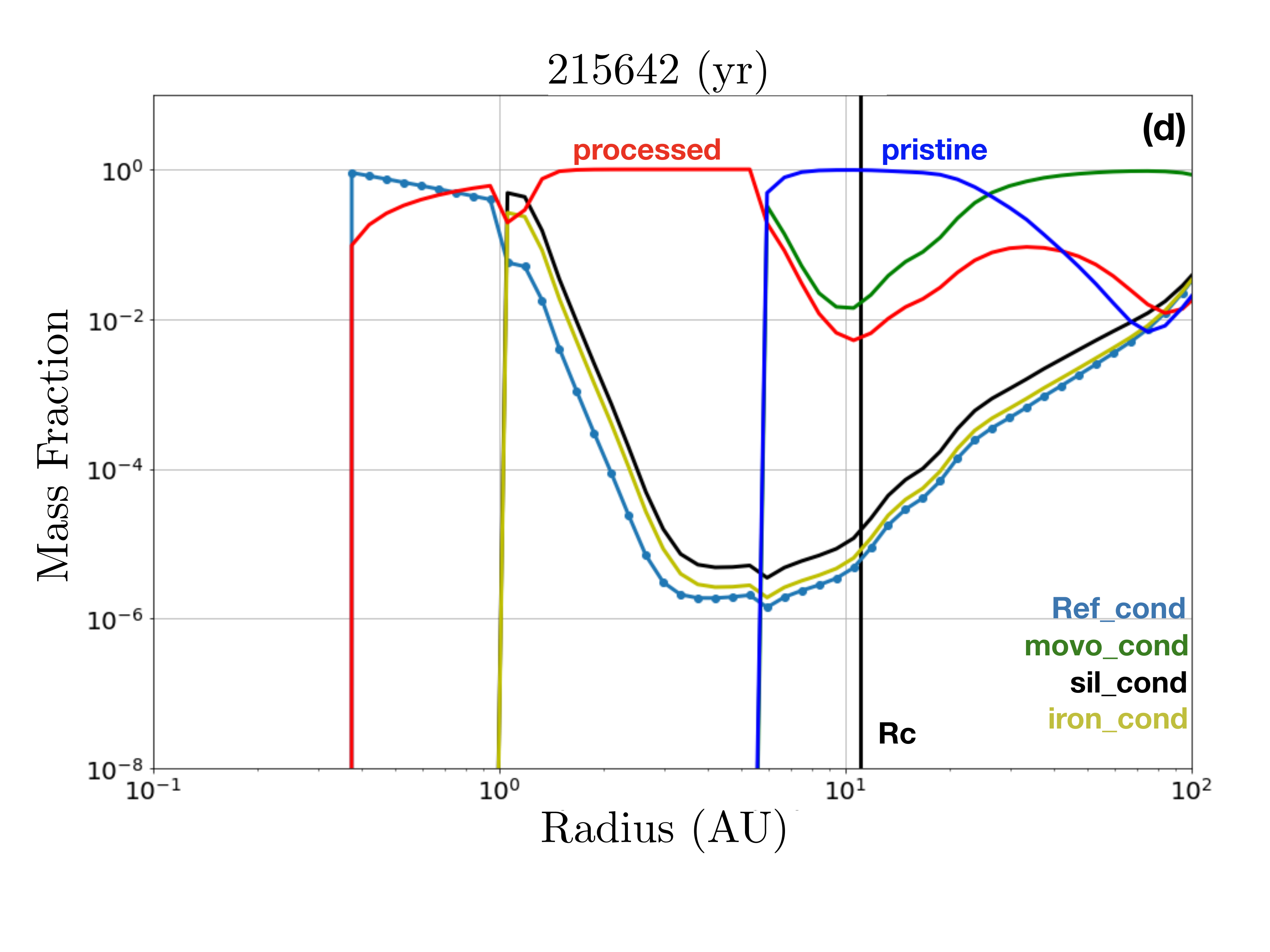}}\\
 \end{center}
\caption{(a) Time evolution of the mass of the forming star, disk and the star+disk system; (b) time evolution of the location of the considered condensation fronts, centrifugal radius, disk edge and inner and outer edges of the dead zone; (c) surface density and (d) mass fraction of dust-components as a function of the heliocentric distance plotted at the end of the collapse.
} 
\label{diskbuild}
\end{figure}
In Fig.~\ref{diskbuild} we report the resulting simulation for the monotonic function. Figure.~\ref{diskbuild}(a) shows the time evolution of the mass of the star, disk and the star+disk system during the first 1 Myr from the start of the cloud collapse. Fig.~\ref{diskbuild}(b) shows, similarly to \citet{2018ApJ...867L..23P}, the time evolution of the location of the all considered condensation fronts, centrifugal radius, disk edge and inner and outer edges of the dead zone that results from the simulation.  Fig.~\ref{diskbuild}(c) shows the resulting surface densities, at the end of collapse for the refractory condensates (red solid line for the $^{26}$Al, blue solid line for the  refractory from cloud) and refractory bulk (condensates plus processed plus pristine) for the $^{26}$Al and for the refractory from cloud.  Plotted are also all the other bulk rocky material (orange line), the water ice (cyan shaded area), and \ce{H2} (black dotted line). The location of the centrifugal radius is also shown. Fig.~\ref{diskbuild}(d) shows the mass fraction of different components as a function of radius and at the end of collapse. Our results are comparable to those presented in \citet{2018ApJ...867L..23P}; see also \citet{2012M&PS...47...99Y}.

\section{Analytic expressions for the injection functions.}
\label{mathexpression}

Here we report the analytic expression for the injection functions that we use for describing the $^{26}\ce{Al}$ abundance as a function of the dimensionless time $x\equiv t/$(150 kyr). In normalized form, they are:
\begin{itemize}

\item{For $x < 0.15$, $f(x)=0$ for both functions.}

\item{
For $0.15\le x \le 0.33
$
, $f(x)= 10\times(x-0.15)^{0.88}(1-(x-0.15))^4$ (this is 1 at $x=0.33$) for both functions.

}

\item{For $ x >0.33, f(x)=1$ for the monotonic function and $f(x)=1.197 -0.598 x$ for the spike.

}

\end{itemize}

%

\vspace{5mm}

\end{document}